\documentclass[prl,twocolumn,preprintnumbers,amsmath,amssymb]{revtex4-1}

\usepackage{amsmath}
\usepackage{amsfonts}
\usepackage{amssymb}
\usepackage{epsfig}
\usepackage{epstopdf}
\usepackage[usenames,dvipsnames]{color}
\newcommand{\braket}[2]{\mbox{$\langle #1|#2\rangle$}}
\newcommand{\ketbra}[2]{\mbox{$|#1\rangle\langle #2|$}}
\newcommand{\op}[1]{\mbox{\boldmath $\hat{#1}$}}

\newcommand{\ket}[1]{\vert#1\rangle}
\newcommand{\bra}[1]{\langle#1\vert}

\begin{document}

\title{Estimation of a quantum interaction parameter using weak measurement: theory and experiment}
%\title{How to estimate the interaction parameter of a weak quantum measurement}
\author{Holger F. Hofmann$^1$, Michael E. Goggin$^2$, Marcelo P. Almeida$^3$, and Marco Barbieri$^{4,5}$}
\affiliation{$^1$Graduate School of Advanced Science of Matter, Hiroshima University, Kagamiyama 1-3-1, Higashi Hiroshima, and JST, CREST, Sanbancho 5, Chiyoda-ku, Tokyo 102-0075, Japan}
\affiliation{$^2$Physics Department, Truman State University, Kirksville. MO 63501}
\affiliation{$^3$ARC Centre for Engineered Quantum Systems, ARC Centre for Quantum Computation and Communication Technology,
School of Mathematics and Physics, University of Queensland, 4072 Brisbane, QLD, Australia}
\affiliation{$^4$Laboratoire Charles Fabry, Institut d'Optique, CNRS, Universit\'e Paris-Sud, Campus Polytechnique, RD 128, 91127 Palaiseau, France }
\affiliation{$^5$Clarendon Laboratory, Department of Physics, University of Oxford, OX1 3PU, Oxford, UK. \vskip 2mm}

\begin{abstract}
We investigate the estimation of a small interaction parameter from the outcomes of weak quantum measurements implemented by the interaction. The relation of weak values and sensitivity is explained and the different contributions of post-selected results are identified using experimental data. The results show how weak values can be used to control the distribution of input state sensitivity between different post-selected outcomes.
\end{abstract}

\maketitle 

Quantum metrology is concerned with the optimal estimation of a specific parameter from a measurement. The maximum achievable precision by which the parameter can be estimated depends on both the initial quantum state of the system and the measurement strategy employed \cite{DAriano01,Giovannetti06,Paris08,Esc11}. The importance of a careful design is particularly relevant when one has to deal with very weak interactions, where a clumsy choice may result in the inability of obtaining meaningful results. In quantum mechanics, such weak interactions can also be used to perform statistical measurements with negligible back-action, so-called weak measurements \cite{Aharonov88, Pryde04}. While limited information can be collected in a single instance of a weak measurement, an estimation of the observable involved can neverthless be obtained over many runs. Remarkably, weak measurements in conjunction with post-selection based on a further ordinary measurement may give rise to measured values outside the spectrum of the observable, as experimentally verified in \cite{Pryde05, Iinuma11}. The appearence of such anomalous values has been put in relation with the failure of a macroscopic realistic model in explaining the  dynamics of such experiments \cite{Williams08, Goggin11}, and to time-symmetric formulations of quantum mechanics \cite{Aharonov10}.  

Since anomalous weak values correspond to meter shifts that are much larger than the ones expected from the measurement interaction, weak measurements may be useful in the determination of small interaction parameters \cite{Hosten08, Dixon09, Starling10}. The possible role of weak values in optical interferometry has recently been addressed \cite{Brunner09}. However, the fundamental relation between the sensitivity of parameter estimates and the observation of weak values is still somewhat unclear \cite{Hofmann11}. Here, we investigate the role of weak values in the estimate of the small interaction parameter used in the measurement. Significantly, we find that the total sensitivity is equal for all post-selection strategies, resulting in maximal sensitivities for a wide range of different output measurements. This result seems to indicate that the weak values provide an error-free evaluation of the obsrevable acting on the meter system under the appropriate post-selection conditions. 

Let us consider the elements involved in weak quantum measurements. A quantum system described by a quantum state $\ket{\psi}$ interacts weakly with a well known probe state. As a result of the interaction, a measurement on the probe leads to some information about the physical property $\op{\mathcal{A}}$ of the system through which the system interacts with the probe. Hence, the probability of each outcome $m$ for the system will depend on the value of $\op{\mathcal{A}}$. As for any quantum measurement, the interaction between system and probe also induces an uncontrollable disturbance in the state of the system. However, this disturbance is negligibly small if the interaction is very weak. It is therefore possible to define the state of the system more precisely by postselecting the result of an additional output measurement $\ket{f}$. This double definition of the quantum state by preparation and postselection results in the weak values $\langle \op{\mathcal{A}}\rangle_{\mathrm{wv}}$, which can be far outside the range of the eigenvalues observed in strong measurements.

The effects of the weak measurement interaction on the probabilities of the measurement outcomes $m$ is determined by the product of a small interaction parameter $\epsilon$ with the weak value $\langle \op{A}\rangle_{\mathrm{wv}}$. To estimate the value of an unknown interaction parameter $\epsilon$, we use a known combination of input state and output measurement. The estimation strategy is thus defined in terms of the initial and final meter states that define the weak values. For the theoretical analysis of the possible preparation and post-selection strategies, we describe the outcomes $m$ of the weak measurements in terms of the measurement operators  $\op{E}_m$ acting on the state of the system. Since we are only interested in the regime of weak interactions, we use a linearized expression: 
\begin{equation}
\label{operator}
\op{E}_m=\sqrt{w_m}\left(\op{I}+\epsilon \kappa_m \op{\mathcal{A}}\right).
\end{equation}
Here, $\kappa_m$ describes the effect of $\op{\mathcal{A}}$ on the probability of the specific outcome $m$ (positive $\kappa_m$ for increase, negative  $\kappa_m$ for decrease), and $w_m$ is the probability of $m$ without the interaction. Note that Eq.(\ref{operator}) defines the interaction parameter $\epsilon$ in terms of the effects of $\op{\mathcal{A}}$ on the measurement statistics. This definition is motivated by the formal similarity to a weak unitary transformation, which is obtained when $\epsilon$ is replaced by an imaginary phase parameter. As pointed out in Ref.\cite{Hofmann11}, phase estimation in quantum interferometry is then equivalent to an estimation of the interaction parameter from imaginary weak values.

The results of a weak measurement are given by the joint probabilities $p(m,f)$ of obtaining a weak measurement result $m$ and observing an output state $\{\ket{f}\}$ in a subsequent projective measurement on the system. According to quantum mechanics, this probability is given by $|\bra{f}\op{E}_m\ket{\psi}|^2$, which includes the measurement back-action as a quadratic term in $\epsilon$ and in $\op{\mathcal{A}}$. For sufficiently weak interactions, this term can be neglected and the joint probabilities of the weak measurement are given by
\begin{equation}
p(m,f)=w_m|\braket{f}{\psi}|^2\left(1+2\epsilon \kappa_m \text{Re}\left(\frac{\bra{f}\op{\mathcal{A}}\ket{\psi}}{\braket{f}{\psi}}\right)\right).
\label{postprob}
\end{equation} 
In this expression, the effects of $\op{\mathcal{A}}$ on the quantum statistics is given by the weak value of $\op{\mathcal{A}}$ obtained by postselection of a final state $\ket{f}$. Eq.(\ref{postprob}) thus confirms that the sensitivity of the output probabilities $p(m,f)$ to the interaction parameter $\epsilon$ is given by the weak values of $\op{\mathcal{A}}$ defined by the choice of input state and output measurement. 

In the formalism of quantum metrology \cite{Helstrom76, Paris08}, the linear dependence of the output probabilities $p(m,f)$ on small changes in the parameter $\epsilon$ is quantified by the logarithmic derivative, 
\begin{equation}
\label{logderiv}
\partial_\epsilon\left.\ln \left( p(m,f) \right)\right|_{\epsilon{=}0}=2 \kappa_m\text{Re}\left(\frac{\bra{f}\op{\mathcal{A}}\ket{\psi}}{\braket{f}{\psi}}\right).
\end{equation}
Thus, the weak values of $\op{\mathcal{A}}$ provide a direct quantitative expression of the sensitivity of the output probabilities $p(m,f)$ to small changes in the interaction parameter $\epsilon$. Using the established procedures of parameter estimation, it is possible to achieve the maximal sensitivity defined by the Cramer-Rao bound \cite{Helstrom76}. The sensitivity can then be given by the Fisher information $F$, which is the inverse of the minimal  
estimation error $\sigma^2_\epsilon$ \cite{Helstrom76}. By normalizing the values of $\kappa _m$ to 
$\sum_mw_m\kappa_m^2{=}1$, we obtain
\begin{equation}
\begin{aligned}
\label{Fisher}
F&=\sum_{m,f}p(m,f)\left(\partial_\epsilon\left.\ln(p(m,f))\right|_{\epsilon{=}0}\right)^2\\
&=4\sum_f p(f) \; \text{Re}\left(\frac{\bra{f}\op{\mathcal{A}}\ket{\psi}}{\braket{f}{\psi}}\right)^2.
\end{aligned}
\end{equation}
Since the sensitivity is given by the average of the squared real parts of the weak values obtained for different postselected outcomes $f$, optimal results are obtained for final measurements with completely real weak values. In this case, $p(f)=|\braket{f}{\psi}|^2$ can be eliminated and the summation results in $F=4\bra{\psi}\op{\mathcal{A}}^2\ket{\psi}$. Therefore, the maximal sensitivity is determined by the average value of $\op{\mathcal{A}}^2$ in the input state, and different measurement strategies for the postselection of $f$ merely result in different distributions of the weak values. Anomalous weak values much larger than the maximal eigenvalues of the observables do contribute more to the sensitivity, but this effect is compensated by the relatively low probability $p(f)$ of such outcomes.

We can now apply the principles explained above to an experiment. The weak measurement was realized using a two-photon controlled-sign (\textsc{c-s}) gate constituted by a single partially polarizing beam splitter ( PPBS) with transmittivity $\eta_V{=}1/\sqrt{3}$  ($\eta_H{=}1$) for the vertical, $V$, (horizontal, $H$) polarization \cite{Langford05,Kiesel05,Okamoto05}. The operation of the gate is such that it introduces a $\pi$ phase shift only to the $\ket{V,V}$ component of the quantum state, i.e. where both input photons are vertically polarized. Input photons were produced by parametric down conversion (PDC) in a bismuth borate nonlinear crystal. The pump beam is a frequency-doubled, pulsed Ti:Sa laser laser ($\lambda{=}820$nm, $\Delta t{=}100$fs, repetition rate 82 MHz, average power $P{=}50$mW). At the two inputs of the gate, one photon is used as the test system, $\mathrm{s}$, and it is prepared in a generic linear-polarization state
$\ket{\mathrm{s}}{=}\cos{\frac{\theta}{2}}\ket{H}_\mathrm{s}{+}\sin{\frac{\theta}{2}}\ket{V}_\mathrm{s}$. The second photon acts as a probe, $\mathrm{p}$, of the effective interaction strength.

A weak measurement of the Stokes parameter ${\op{S}_{HV}{=}\ketbra{H}{H}-\ketbra{V}{V}}$ can be realized by setting the input polarization of the probe photon to $\ket{H}_\mathrm{p}{+}\epsilon\ket{V}_\mathrm{p}$  \cite{Pryde04, Pryde05, Goggin11}. If the system photon is horizontally polarized, the probe photon remains slightly biased towards the diagonal polarization $D$. However, an interaction with a vertically polarized system photon changes the bias to the opposite diagonal polarization $A$. Therefore, a measurement of the diagonal output polarization realizes the measurement operation \cite{Pryde04,Pryde05} given by Eq.\eqref{operator}, with $\op{\mathcal{A}}=\op{S}_{HV}$, ${\kappa_D={-}\kappa_A=1}$, and ${w_D=w_A=1/2}$.

To realize the parameter estimation, a test interaction parameter was selected by using the corresponding input polarization for the $\mathrm{p}$-photon. The estimation strategy can then be defined by any combination of input state and output measurements. In the present experiment, we used a fixed post-selection measurement on the $\mathrm{s}$-photon and varied the polarization of the system input state to obtain different weak values. Specifically, the post-selection of the $\mathrm{s}$ photon was performed by a destructive measurement in the diagonal basis (${f}{=}{D}$ or ${f}{=}{A}$). To find the output probabilities, count rates were evaluated for different combinations of diagonal polarizations in the $\mathrm{p}$-photon $(m=D,A)$ and the $\mathrm{s}$-photon $(f=D,A)$. Since the distribution of these count rates should be described by Eq.\eqref{postprob}, it is possible to obtain an experimental estimate of the interaction parameter $\epsilon$ from the data. 

\begin{figure}[t!]
\includegraphics[viewport=70 00 1030 630, clip, angle=0, width=\columnwidth]{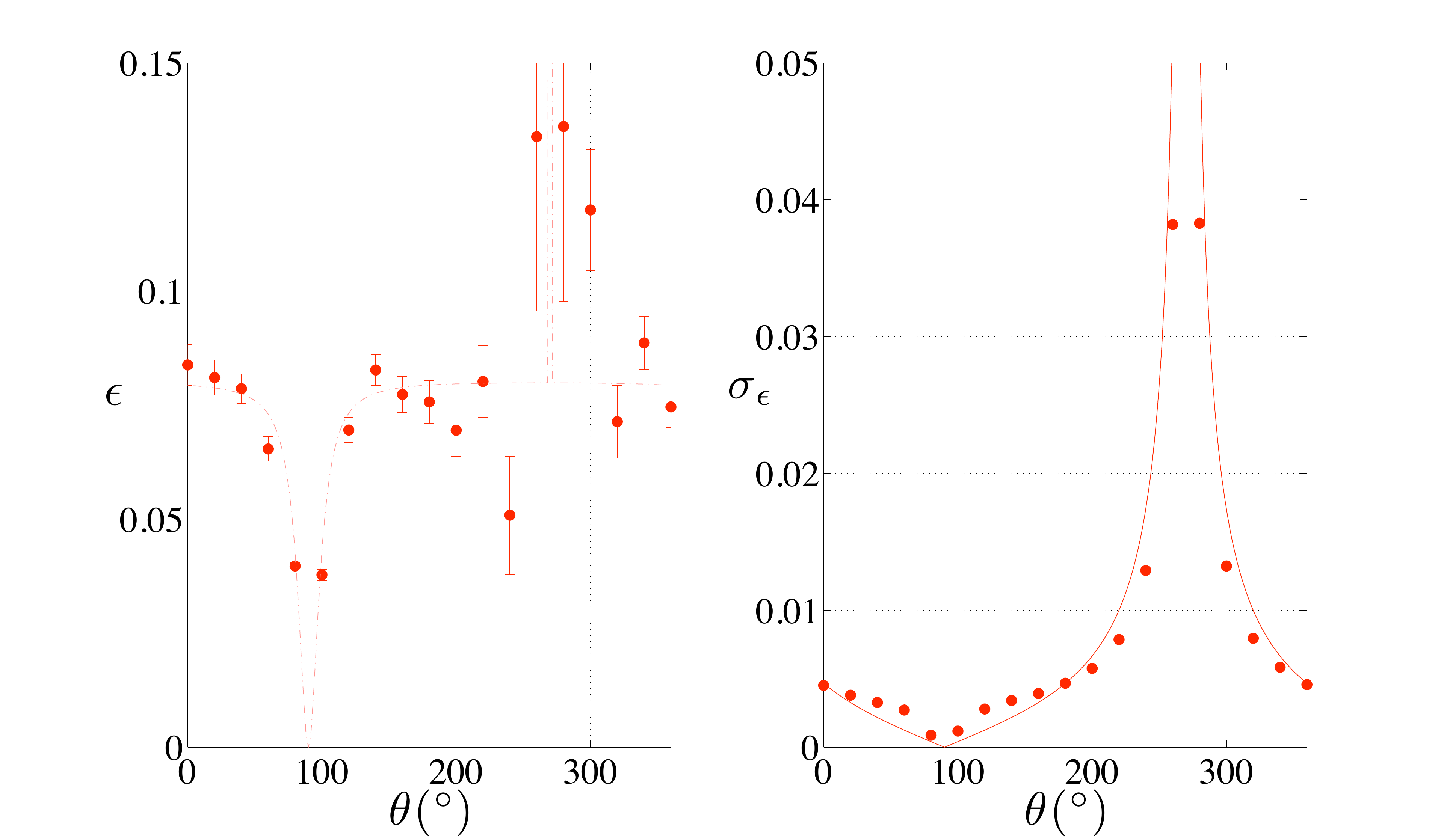}
\caption{Experimental results for the estimate of the interaction parameter using the postselected probabilities with $f{=}A$. Left panel: estimate of the value of $\epsilon$ as a function of the input state $\ket{\mathrm{s}}{=}\cos{\frac{\theta}{2}}\ket{H}{+}\sin{\frac{\theta}{2}}\ket{V}$; the solid line indicates the set value, red dots the experimental data. %%We also show the expected results from a model of our gate as a dotted line. 
Right panel: statistical uncertainties $\sigma_\epsilon$; the points represent the experimental uncertainty due to Poissonian noise on the count rate, the solid line is the theoretical expectation based on Eq. \eqref{postprob}.} 
\label{Epsilon}
\end{figure}

First, we focus on the conditional statistics obtained from a particular post-selection event $f$. According to Eq.(\ref{postprob}), an estimate of the interaction parameter $\epsilon$ can be obtained from the difference between the conditional probabilities $p(D|f)$ and $p(A|f)$ for the two weak measurement outcomes,
\begin{equation}
\label{estimate}
\epsilon=\frac{p(D|f)-p(A|f)}{2 \langle \op{S}_{HV}\rangle_{wv}},
\end{equation}
where $ \langle \op{S}_{HV}\rangle_{wv}$ is the theoretical value of the weak value in the limit of no interaction. The statistical error for this estimate is given by the binomial distribution of outcomes between $m=A$ and $m=D$. We have evaluated this error and confirmed that its inverse corresponds to the contribution to the total Fisher information from the outcome $f$ in Eq. \eqref{Fisher}. The results of this analysis are shown in Fig. \ref{Epsilon}. The left panel shows the estimates obtained at different input polarizations. Significant discrepancies can be observed around $\theta{=}90^\circ$ ($\ket{\mathrm{s}}{=}\ket{D}$), where the breakdown of the linear approximation used in Eq. \eqref{estimate} results in a reduction of the effective weak value \cite{Iinuma11}, and around $\theta{=}270^\circ (\ket{\mathrm{s}}{=}\ket{A})$, where the low sensitivity of the estimate also results in an amplification of errors due to experimental imperfections of the setup. The right panel shows the statistical errors of the estimates, derived from the binomial statistics of the two possible outcomes. Significantly, these uncertainties follow the trend described by Eq. \eqref{postprob}, resulting in particularly low errors around input polarizations of $\theta{=}90^\circ$, orthogonal to the post-selected $A$-polarization. These values are nevertheless affected by a systematic shift inherent to the validity of the approximation in Eq.\eqref{operator}: this sets the limit of applicability of our treatment. 

%%% Figure 2 %%%%%
%%%

\begin{figure}[t]
\includegraphics[viewport=0 0 900 630, clip, angle=0, width=\columnwidth]{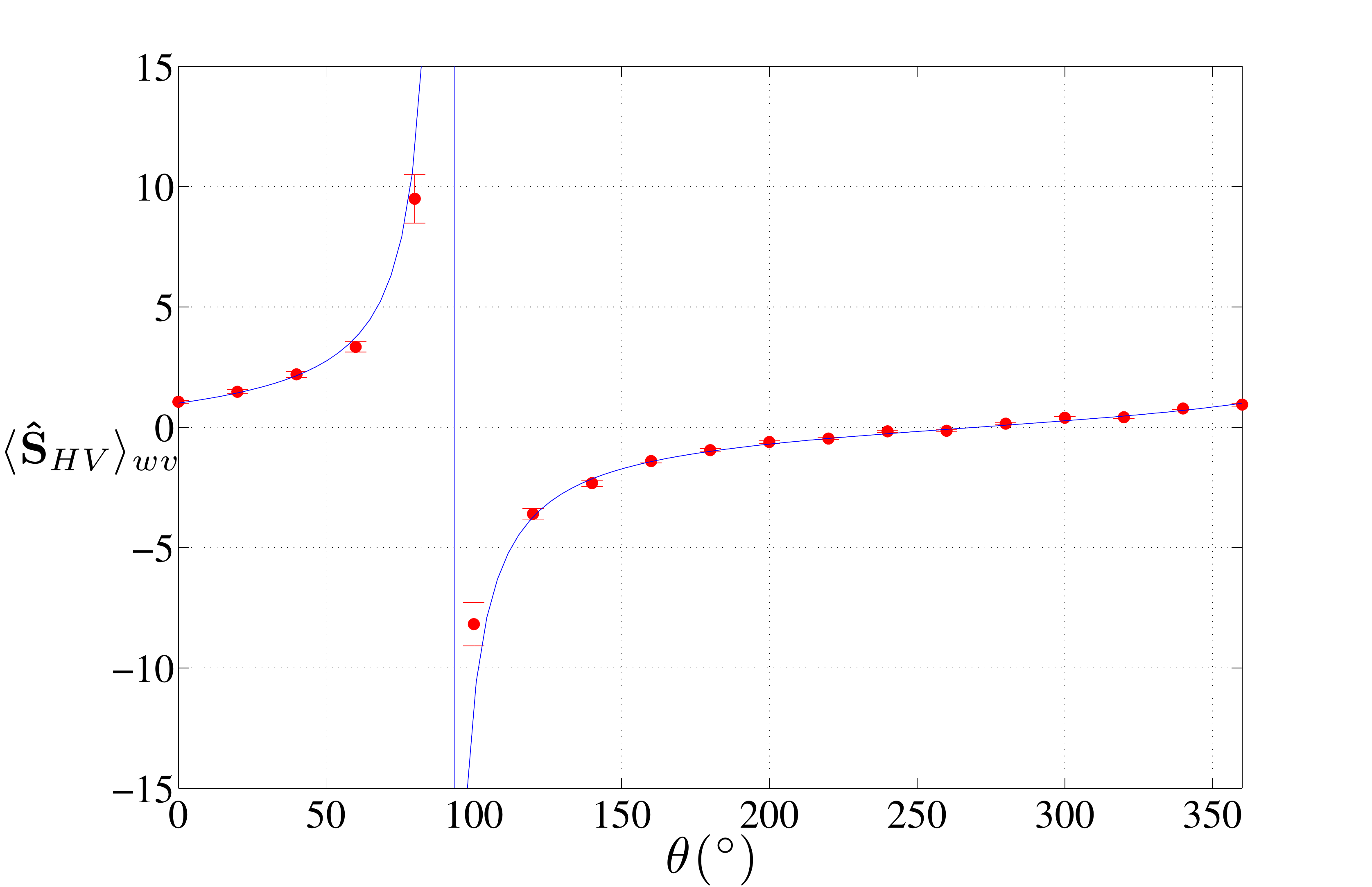}
\caption{Evaluation of the weak values from the logarithmic derivatives of the output probabilities as a function of input state polarization. These quantify the sensitivity of the response of our apparatus to a change in the interaction strength $\epsilon$. Here, $\theta$ is the orientation of polarization on the Poincar\'e sphere, where $\theta=0$ corresponds to horizontal polarization and $\theta=180^\circ$ corresponds to vertical polarization. The data was obtained for a postselection of the anti-diagonal output polarization $A$, corresponding to an angle of $\Theta=270^\circ$. Red dots show the results obtained from the experimental data with error bars representing Poissonian statistical errors. The solid line shows the predicted weak values.} 
\label{WeakLog} 
\end{figure}

According to Eq. \eqref{logderiv}, the sensitivity achieved by a statistical estimate of $\epsilon$ can be explained in terms of the weak values associated with the input polarization $\theta$ and the final polarization measurement $f$. Specifically, the weak values for a final polarization measurement $f$ can be determined from the change in output probabilities caused by differential changes in the interaction parameter $\epsilon$,
%\begin{equation}
%\label{weakNew}
%\langle \op{S}_{HV}\rangle_{wv}= \frac{1}{2} \frac{\partial}{\partial\epsilon}\ln(p(D|f))= - \frac{1}{2} \frac{\partial}{\partial\epsilon}\ln(p(A|f)).
%\end{equation}
\begin{equation}
\label{weakNew}
\langle \op{S}_{HV}\rangle_{wv}= \frac{1}{2} \partial_\epsilon \ln(p(D|f))= - \frac{1}{2} \partial_\epsilon \ln(p(A|f)).
\end{equation}
We confirmed this relation for a postselection of anti-diagonal polarization $f=A$ on input states with variable linear polarization by approximating the derivative as a finite differential between a low coupling of $\epsilon{=}0.08$ and zero coupling and taking the average of the two values obtained for the two meter outcomes $m$. The results, which are in good agreement with the predicted weak values, are shown in Fig. \ref{WeakLog}. Significantly, the region of low estimation errors and high sensitivity around $\theta{=}90^\circ$ coincides with the rapid increase of anomalous weak values when initial polarization and final polarization are nearly orthogonal. 

%%%% Figure 3 %%%%%
%%%%%%%%%%%%%%

The results presented so far correspond to the contribution of the post-selected outcome $f=A$ to the total sensitivity represented by the Fisher information in Eq. \eqref{Fisher}. Although the increase in sensitivity achieved by the anomalous weak values may appear to be impressive, its impact on the total Fisher information is limited by the small post-selection probability. As explained above, the total sensitivity is given by $F=4\bra{\psi}\op{\mathcal{A}}^2\ket{\psi}$, depending only on the observable $\op{\mathcal{A}}$ and the initial state $\ket{\psi}$. In the case of the Stokes parameter $\op{S}_{HV}$, the Fisher information is $4$ for all input states.  

\begin{figure}[t!]
\includegraphics[viewport=70 00 1030 630, clip, angle=0, width=\columnwidth]{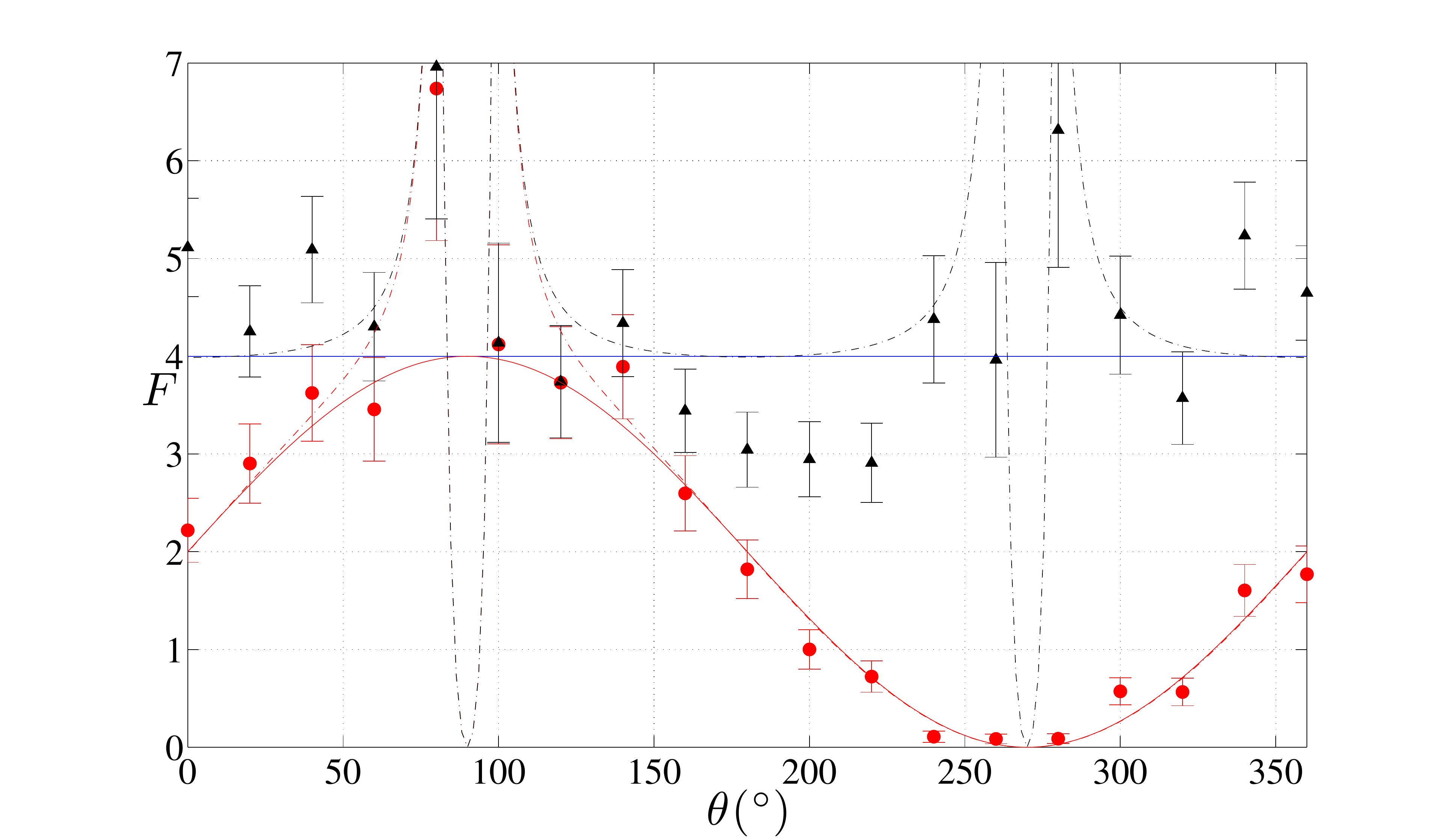}
\caption{Experimental results for the Fisher information. Red dots show the results for the contribution associated only with $\ket{A}_\mathrm{s}$; the red solid line shows the theoretical predictions for the ideal case. Error bars are estimated from the Poissonian statistics of the coincidence counts, resulting in larger uncertainties for anomalous values. The black triangles show the experimental values for the total Fisher information including both post-selection events. Dotted lines illustrate the expected behavior from a model of our gate: this indicates that values $F{>}4$ are merely experimental artifacts. We can attribute the dispersion of the data to the presence of higher-order terms in PDC, and to reflectivities of the PPBS departing from the ideal values, which influence asymmetrically the post-selection on $\ket{A}_s$ and $\ket{D}_s$.} 
\label{FisherInfo}
\end{figure}

In the region around $\theta{=}90^\circ$, where anomalous weak values can be observed, most of the Fisher information originates from the rather low number of postselected events with weak values far greater than the maximal eigenvalues of $\pm 1$. This behavior is in qualitative agreement with the results of Ref. \cite{Starling10}, where they show the advantage in using postselection to avoid problems of saturation and classical noise without losing sensitivity. Our analysis show that such postselection is effectively excluding events which carry little or no information about the interaction parameter $\epsilon$. In order to show this, we plot the Fisher information and the contribution from the postselected state $\ket{A}_s$ as a function of $\theta$ in Fig. \ref{FisherInfo}. The difference between the two plots corresponds to the contribution from the postselected state $\ket{D}_\mathrm{s}$. It should be noted that the agreement between prediction and result is fairly good for the postselected state $\ket{A}_\mathrm{s}$, but much worse for $\ket{D}_\mathrm{s}$. We can attribute this to an intrinsic asymmetry in the operation of the gate in the presence of imperfect reflectivities and noise from higher-photon number contributions, as already observed in \cite{Gillett10}. 

%%%% Figure 4 %%%%%
%%%%

To understand the significance of the result, it is important to consider the relation between statistics and weak values in more detail. While it seems natural that extreme weak values result in higher sensitivities, it is not at all clear how the effective strength of the interaction can depend on the choice of the final measurement $f$. The present results indicate that the sensitivity is proportional to the squared value of $\op{S}_{HV}$, since the effect of a small change in the interaction parameter is proportional to the actual value of $\op{S}_{HV}$. It is therefore possible to interpret the weak value as an estimate of the actual value of $\op{S}_{HV}$ in the quantum fluctuations of $\ket{\psi}$. However, this interpretation seems to highlight the paradoxical nature of weak values: obviously, a straightforward evaluation of the actual value of $\op{S}_{HV}$ is obtained from a measurement of the eigenstates of 
$\op{S}_{HV}$. Since the weak values obtained from other measurements are different from these eigenvalues, one would expect an additional error when weak values are used to estimate the actual value of $\op{S}_{HV}$. However, the present results indicate otherwise. For the purpose of estimating the interaction strength, weak values apparently represent a precise evaluation of the system property in question. Consequently, there is a rather surprising freedom of choice in the selection of the final measurement used to determine the output state of the system. The estimation of interaction parameters using weak measurement thus reveals an amazing flexibility in the way that quantum mechanics distributes the available information between physical properties, with fundamental implications for the way we think about the counterintuitive properties of quantum systems. We have also investigated how this translates into an experiment, revealing that, while the sensitivity follows closely our prediction, systematic effects might affect the behaviour of anomalous values. 

We thank Andrew White, Benjamin Lanyon, Giuseppe Vallone and Luca Pezz\'e for valuable discussions. H.F.H. is supported by the Grant-in-Aid program of the Japanese Society for the Promotion of Science, JSPS. M.P.A. acknowledges the financial support from the ARC Centre of Excellence, Discovery Federation Fellow programs and an IARPA-funded US Army Research Office contract. M.B. is supported by the Marie Curie contract PIEF-GA-2009-236345-PROMETEO. M.E.G. would like to thank the University of Queensland for support and hospitality during his sabbatical visit there.

\end{document}